\documentstyle[aps]{revtex}
\begin{document}
\input{psfig}
\title{Surface spin-flop phases and bulk discommensurations in
antiferromagnets}
\author{C Micheletti\dag, R B Griffiths\ddag\ and J M Yeomans\dag}
\address{\dag\ Theoretical Physics, Oxford University,
1 Keble Road, Oxford OX1 3NP, UK}
\address{\ddag\ Physics Department,
Carnegie-Mellon University, Pittsburgh, PA 15213}

\date{November 1996}
\maketitle
\begin{abstract}

Phase diagrams as a function of anisotropy $D$ and magnetic field $H$
are obtained for discommensurations and surface states for a model
antiferromagnet in which $H$ is parallel to the easy axis. The surface
spin-flop phase exists for all $D$. We show that there is a region
where the penetration length of the surface spin-flop phase
diverges. Introducing a discommensuration of even length then
becomes preferable to reconstructing the surface. The results are used
to clarify and correct previous studies in which discommensurations
have been confused with genuine surface spin-flop states.

\end{abstract}

\pacs{75.10.Hk, 75.50.Ee, 75.30.Gw}

Recent experimental results on Fe/Cr(211) superlattices
\cite{WM94,WM94b} have
stimulated theoretical and numerical work aimed at explaining the
occurrence of series of phase transitions in samples with an even
number of Fe blocks. The antiferromagnetically-coupled Fe blocks can
be described phenomenologically by a chain of classical X-Y spins with
antiferromagnetic interactions and uniaxial spin anisotropy $D$. For a
chain of infinite length, application of an external magnetic field $H$
in a direction parallel to the easy axis (the spin direction in zero
field) results in a transition at a finite field $H=H_{SF}(D)$ to a 
spin-flop phase in which the spins on the two sublattices are
approximately perpendicular to the applied field, and nearly opposite
to each other \cite{Neel}.  

For finite chains the situation is more complex. 
In 1968 Mills \cite{M68} suggested that
spins in a layer near a free surface of such an antiferromagnet could
rotate into the flopped state at a field $H'_{SF}$ significantly less
than $H_{SF}$. To understand this further we present numerical and
analytic studies for an antiferromagnetic chain  
for all values of $H$ and $D$. We show that a consideration of
{\em both} surface states {\em and} discommensurations is necessary
to fully understand the phase diagram. 
It transpires that discommensurations, while they are not surface
phenomena {\em per se} can provide the minimum energy configuration for both
semi-infinite and finite systems. The results allow us to unify certain
aspects of previous work \cite{WM94,WM94b,M68,T94,KC,T95} 
while correcting others.

	Consider a chain of classical spins described by a Hamiltonian
\begin{equation}
{\cal H}  = \sum_i \biggl[ \cos(\theta_i - \theta_{i+1}) - H
\cos \theta_i -
{D \over 4} \cos (2 \theta_i) \biggr],
\label{e.ham}
\end{equation}
where the exchange coefficient has been taken as the unit of energy,
$\theta_i$ is the angle between the $i$th spin and the direction
of the field $H$, and $D$ is a two-fold spin anisotropy. One can
think of $\theta_i$ as the direction of the magnetization in the $i$th
layer of an antiferromagnet, or in the $i$th  Fe layer in a
Fe/Cr superlattice. It is appropriate to work at zero temperature in
one dimension because in-plane fluctuations are not important for the
superlattice physics.

\begin{figure} 
\centerline{\psfig{figure=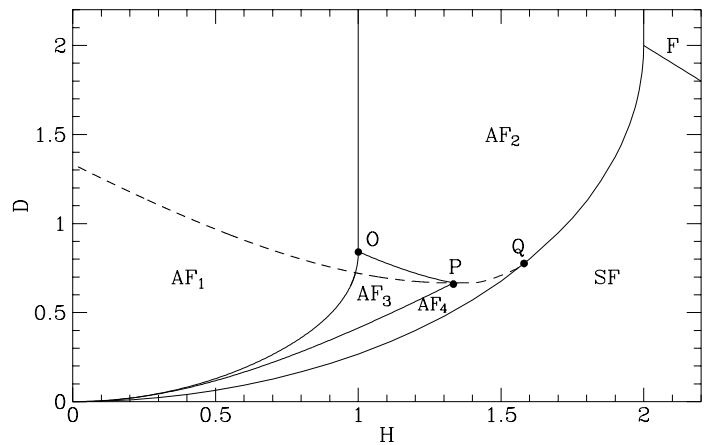,width=4.0in}}
\caption{Phase diagram in
the $H,D$ plane.  Additional details are in Fig.~\protect{\ref{f2}}.}
\label{f1}
\end{figure}

For a bulk system, in which the sum in (\ref{e.ham}) runs from
$-\infty$ to $+\infty$, the ground state is either ferromagnetic (F)
with all spins parallel to the field, antiferromagnetic (AF) with the
spins alternating between $0$ and $\pi$, parallel and antiparallel to
the field, or a spin flop (SF) phase in which the spins alternate
between $+\phi$ and $-\phi$, for some $\phi$ in the range $0 < \phi <
\pi/2$. The regions where these different phases are stable are
indicated in Fig.~\ref{f1}, where the AF phase is above the line
$D(4-D)=H^2$ and to the left of $H=2$, while the F and SF phases are
separated by the line $D+H=4$ \cite{AC}.  The additional structure in
the AF region refers to discommensurations and surface phases.

We first describe the phase diagram of a chain which is constrained by
suitable boundary conditions to include a discommensuration \cite{n1}.
Note that the bulk AF ground state is degenerate with two possibilities:
$\theta_i$ equal to $0$ for $i$ even and $\pi$ for $i$ odd, or {\em
vice versa}. A discommensuration is a boundary or interface between
two regions corresponding to these two possibilities; in particular, a
configuration in which $\theta_{2n}$ tends to $0$ and $\theta_{2n+1}$
to $\pi$ as $n$ tends to $+\infty$, while $\theta_{2n}$ tends to $\pi$
and $\theta_{2n+1}$ to $0$ as $n$ tends to $-\infty$. For all values
of $D$ above the dashed line in Fig.~\ref{f1}, the discommensuration
of minimum energy is of the Ising type, $\ldots 0,\pi,0,\pi,0,0,\pi,0,
\pi,0 \ldots $ , with two adjacent spins parallel to each other and to
the field, in the middle of what is otherwise an antiferromagnetic
configuration.

\begin{figure}
\centerline{\psfig{figure=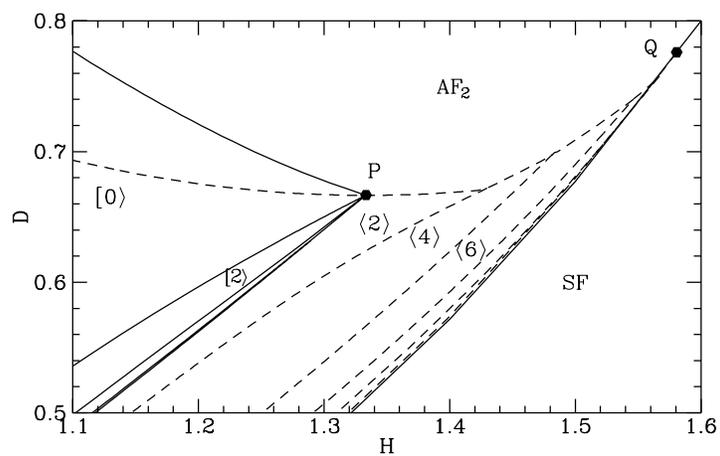,width=4.0in}}
\caption{Phase diagram: details near the points $P$ and $Q$.}
\label{f2}
\end{figure}

        In the AF region below the dashed line in Fig.~\ref{f1},
``flopped'' discommensurations of different length have lower energies
than the Ising discommensuration.  A flopped
discommensuration of type $\langle 2m\rangle$ consists of a ``core''
of $2m$ spins in which the spin configuration resembles that in a bulk
SF phase, located between ``tails'', each of which rapidly reverts to
the configuration of the corresponding AF phase with increasing
distance from the core. See, for example, Fig.~1 in \cite{Pap}.
As $D$ decreases, the distinction between the ``tails'' and the ``core''
is less clear, but we continue to use the same label for the
discommensuration which evolves continuously from $\langle 2m\rangle$
at larger $D$.  One can think of such a discommensuration as composed
of a pair of AF--SF and SF--AF interfaces centered at the points where
the corresponding tail joins the core.  At low values of $H$, the
discommensuration $\langle 2 \rangle$ has the lowest energy, but upon
approaching the bulk AF:SF phase boundary, one finds---see
Fig.~\ref{f2} for details omitted from Fig.~\ref{f1}---a sequence of
phase transitions to $\langle 4 \rangle$, $\langle 6 \rangle$, \dots \
as $H$ increases.  Our numerical procedures, which used effective
potential techniques \cite{CG,FG,KH} to construct surface states by
extrapolating from the bulk \cite{Mth}, 
found values of $2m$ up to 14. We were able to
trace the first-order lines separating the different $\langle 2m
\rangle$ phases down to a value of $D$ between 0.1 and 0.4.  We found
no evidence to suggest that these lines end in critical points, and we
believe it likely that they will persist all the way down to $D=0$.

	These transitions reflect the discrete nature of the spin
chain. Thus it is not surprising that they are absent in the continuum
approximation employed in \cite{Pap} for small values of $D$. That
study showed that the width of the discommensuration tends to $\infty$
at the AF:SF boundary, in agreement with what we found for larger
values of $D$.

	The boundary between $\langle 2 \rangle$ and the Ising
discommensuration is second order, and represents the limit of
stability of the latter as $D$ decreases. An analytic calculation
yields the equation $(D+H -1)^{-1} = 5/3 + D -H$, in good agreement
with our numerical calculations, and those in \cite{Pap} when
$H=0$. The boundaries between the Ising discommensuration and $\langle
4 \rangle$, $\langle 6\rangle$ etc., are first order, and were
obtained numerically.  The triple points at which the phases AF$_2$,
$\langle 2m \rangle$ and $\langle 2m+2 \rangle$ meet tend to an
accumulation point, $Q$, located at $H\approx 1.58, D\approx 0.78$.
Presumably this is where the energy to create a pair of AF--SF and
SF--AF interfaces (infinitely far apart) is equal to the energy of an
Ising discommensuration.


	We now consider the surface states of a semi-infinite chain.
If a ground-state configuration of an infinite chain is cut in two
while the spins are held fixed, there results two ideal or {\em
unreconstructed} surfaces.  Sometimes the energy of the surface state
can decrease through a local reconstruction in which spins near the
surface are altered from their bulk values by amounts which tend
asymptotically to zero with increasing distance from the surface.  The
energy {\em change} during such a reconstruction can be calculated
from the appropriate Hamiltonian, which is (\ref{e.ham}) with the sum
going from $i=0$ to $+\infty$, even though the total energy is not
well defined.  Hence the reconstructed surface of minimum energy is
(usually) well-defined.

	While the surface reconstruction in the spin-flop phase occurs
smoothly, so that there are no phase transitions, the situation in the case
of antiferromagnetic surface states is more complicated. There are two
types of surface states, $A$ and $B$, with unreconstructed versions
having the surface spin parallel ($\theta_0=0$) or opposite
($\theta_0=\pi$) to the field direction respectively.
Even after reconstruction takes place, so that $\theta_0$ has
  changed, the $A$-type ($B$-type) surface can still be identified
  through the fact that $\theta_{2n}$ tends to $0$ ($\pi$) as
  $n\rightarrow\infty$.

	Throughout the AF region, $A$-type surfaces do not
reconstruct.  The
behavior of $B$-type surfaces is more complicated.  In region AF$_1$ in
Fig.~\ref{f1}, the unreconstructed surface has the lowest energy.  In
region AF$_2$, which meets AF$_1$  along a line $H=1$ for arbitrarily
large $D$, there is a set of degenerate (equal minimum energy)
reconstructed surfaces,
\begin{equation}
[ 0 \rangle =0\, 0\, \pi\, 0\, \pi\, 0\, \pi \ldots,\quad
[ 2 \rangle =0\, \pi\, 0\, 0\, \pi\, 0\, \pi \ldots,
\label{e.surf}
\end{equation}
where $[2n\rangle$ consists of $2n$ spins in an antiferromagnetic
arrangement, followed by two spins parallel to the field, and then the
bulk antiferromagnetic phase.  One can think of this reconstructed
surface as an Ising discommensuration located a certain distance from
the surface.  Because the ``tails'' of this discommensuration have
zero length, it does not interact with the surface, and its energy is
independent of its distance from the surface as long as
$\theta_0=0$. (This artificial degeneracy would be lifted if weak
longer-range interactions were included in (\ref{e.ham}).)

	In the AF$_3$ region in Fig.~\ref{f1}, the $B$-type surface again
reconstructs, but the spins near the surface are no longer locked in Ising
positions.  Just as in the AF$_2$ region, one can think of the surface state as
consisting of a discommensuration located a finite distance from the surface,
but now this discommensuration is of the flopped type with a core of length
two, and tails extending out on either side of the core.  Because of
the tails, the
discommensuration interacts with the surface, and the minimum surface energy
occurs when it is some specific distance $2n$ from the
surface, depending on $D$ and $H$.  Thus in AF$_3$ one finds genuine spin-flop
surface states.  As $H$ increases, the discommensuration moves further from the
surface. It does this, at least when $D$ is large,
discontinuously in steps of 2, via a series of first-order phase
transitions, some of which are shown in Fig.~\ref{f2}, where they
extend leftwards
from the point $P$.  Numerically we have seen states with $2n$ up to 14, and
our results are consistent with $n$ tending to infinity at the right side of
the AF$_3$ region, which our analytic calculations, in agreement with
\cite{T94}, show to be the line
\begin{equation}
 D=\sqrt{1+H^2} -1.
\label{e.bound1}
\end{equation}
The upper boundary of the AF$_3$ region extending from $O$ to $P$ is a
continuous (second-order) transition representing the limit of stability of the
Ising surface phase $[0\rangle$, see (\ref{e.surf}), as $D$ decreases. An
analytic
calculation yields the implicit equation
\begin{equation}
( 2 + D - H -1/a)^{-1}= 2 + D +H -a,
\label{e.bound2}
\end{equation}
where $a= H + D + 1/(1 -H -D)$.  Thus the point $P$, where all the phases
$[2n\rangle$ come together, lies at $H=4/3,$ $D=2/3$, the intersection of
(\ref{e.bound1}) and (\ref{e.bound2}).  Both (\ref{e.bound1}) and
(\ref{e.bound2}) agree with our numerical results.

	We find that the first-order lines extending downwards and
leftwards from $P$ in Fig.~\ref{f2}, separating the canted phases
$[2n\rangle$ from $[2n+2\rangle$, end in critical points as $D$
decreases. The larger the value of $n$, the further the line extends
towards the origin, but presumably for any finite value of $n$ the
difference between the phases $[2n\rangle$ and $[2n+2\rangle$
eventually disappears at some finite value of $D$. Because this value
decreases with increasing $n$, it is plausible that the corresponding
critical points accumulate at the origin.

	As is evident in Fig.~\ref{f1}, the region AF$_3$ becomes extremely
narrow as $D$ decreases. The left boundary
approaches a parabola $D=0.5 H^2$ to within numerical precision, which is
asymptotically the same as (\ref{e.bound1}).  We nonetheless believe that the
width of AF$_3$ remains finite as long as $D>0$.
Numerical evidence for this is that the value of the surface spin
$\theta_0$, at the left edge of the $AF_3$ region
(that is for $H$ just large enough to produce the surface spin-flop
phase) tends to a value near $\pi/3$ as $D$ goes to zero.
This indicates that even for very
small $D$ the discommensuration at the threshold field is still a finite
distance from the surface.

	Between AF$_3$ and the AF:SF bulk phase boundary lies region AF$_4$ in
which the flopped discommensuration is repelled by the surface, so that its
minimum energy location is in the bulk infinitely far away from the
surface, as noted in ref. \cite{T94}. Thus there is not
a minimum-energy reconstructed $B$ surface, or, properly speaking, a ``surface
spin-flop phase'' in region AF$_4$.  It would probably be better to identify
AF$_4$, thought of as part of the $B$-type surface phase diagram, as the
``discommensuration phase'', since the minimum energy surface will always be
of the $A$-type.

         In retrospect it seems likely that the broadening of the SSF
transition mentioned in the abstract of ref. \cite{KC} actually
refers to broadening of the bulk discommensuration which, as noted
above, occurs as $H$ approaches the AF:SF boundary inside region
AF$_4$.  It seems that no earlier work has correctly identified the
stable SSF phase at small values of $D$, characterized when it first
appears with increasing $H$ by a surface spin with a value very near
$\pi/3$.  The narrowness of the AF$_3$ region for
small $D$ may be why it was overlooked.

	For a finite chain of length $L$, the minimum energy is achieved with
$A$-type surfaces at both ends when $L$ is odd.  This is not possible if $L$ is
even, so when $H$ is small there will be an $A$-type surface at one end and a
$B$-type at the other.  As $H$ increases, the latter reconstructs through a
discommensuration which then moves towards the centre, where it arrives when
the value of $H$ is approximately that given by (\ref{e.bound1}).  Next the
discommensuration proceeds to broaden in discontinuous steps, with the AF-SF
and SF-AF interfaces on the left and right side of the core moving outwards,
until they arrive at the two surfaces when $H$ is approximately at the bulk
AF:SF transition value.  The numerical results reported in
\cite{WM94,WM94b,T95} for even values of $L\geq 16$ and $D=0.5$ agree well with
the phase diagrams in Figs.~\ref{f1} and \ref{f2}.  The same broadening
scenario for the discommensuration, at a much smaller value of $D$, is visible
in Fig 3 of \cite{T94}.

JMY and CM acknowledge support from the EPSRC.  We thank D. L. Mills
for bringing \cite{Pap} to our attention.

\end{document}